\documentclass{Interspeech2024}
\usepackage[utf8]{inputenc}

\usepackage{amsmath}
\usepackage{hyperref}
\usepackage{url}
\usepackage{graphicx}

\usepackage{lipsum}

\usepackage{pifont}
\newcommand{\cmark}{\textcolor{teal}{\ding{51}}}%
\newcommand{\xmark}{\textcolor{RawSienna}{\ding{55}}}%
\usepackage{float}
\usepackage[dvipsnames]{xcolor}
\usepackage{subcaption}
\usepackage{booktabs,caption}
\usepackage[flushleft]{threeparttable}
\usepackage{enumitem}
\setlist[enumerate]{leftmargin=*, topsep=3pt}

\interspeechcameraready

\title{STraDa: A Singer Traits Dataset}

\name[affiliation={1, 2}]{Yuexuan}{Kong}
\name[affiliation={1}]{Viet-Anh}{Tran}
\name[affiliation={1}]{Romain}{Hennequin}

\address{
  $^1$Deezer, France\\
  $^2$Nantes Université, École Centrale Nantes, CNRS, LS2N, UMR 6004}
\email{ykong@deezer.com}

\keywords{large-scale dataset, singer metadata, singer traits, singing voice analysis, singer sex classification}

\begin{document}

\maketitle

\begin{abstract}
There is a limited amount of large-scale public datasets that contain downloadable music audio files and rich lead singer metadata. To provide such a dataset to benefit research in singing voices, we created Singer Traits Dataset (STraDa) with two subsets\footnote{All data and description can be found at: \url{https://zenodo.org/records/10057434}}: automatic-strada and annotated-strada. The automatic-strada contains twenty-five thousand tracks across numerous genres and languages of more than five thousand unique lead singers, which includes cross-validated lead singer metadata as well as other track metadata. The annotated-strada consists of two hundred tracks that are balanced in terms of 2 genders, 5 languages, and 4 age groups. To show its use for model training and bias analysis thanks to its metadata's richness and downloadable audio files, we benchmarked singer sex classification (SSC) and conducted bias analysis.

\end{abstract}

\section{Introduction}
In the expansive landscape of machine learning, the role of high-quality datasets cannot be overstated. In the field of studies on singing voices, downloadable audio segments, accompanied by rich metadata, serve as the foundation upon which advanced models and analytical methodologies are built. However, despite the high interest, the availability of large-scale datasets focusing on singing voices remains notably limited\cite{ramona2008vocal, bayle2016large, chowdhury2020jukebox}. Moreover, the annotations regarding the lead singer are often unclear due to difficulties in collecting annotations. 

To provide the community with a dataset with downloadable audios and cross-validated annotations, we created a singer traits dataset STraDa that has two subsets. One subset (referred as automatic-strada) is created automatically by matching sources and larger in scale , and a subset smaller in size (referred as annotated-strada) which is carefully curated and annotated manually to ensure subgroup balance and accuracy. 

The automatic-strada comprises 169 hours of audio, consisting of 25194 excerpts of 30s, each from a different song, performed by 5264 singers. We provided unique identification codes for researchers to download the audio excerpts from Deezer database for their own research purposes. There are 3426 male singers, 1827 female singers, and 11 singers identified with non-binary genders. Tracks in automatic-strada are diverse, spanning 25 genres and 35 languages, and were recorded under a wide range of conditions. Table \ref{genre} shows the number of tracks that certain subgroups contain. To minimize the potential for errors in annotation, automatic-strada metadata was cross-validated across four different sources. Another important aspect of automatic-strada is that, in order to facilitate pre-processing and reduce noise in the annotations, only tracks featuring a single lead singer were chosen, an unique aspect not found in previous datasets.

\begin{table}[H]
\centering
\begin{subtable}{\linewidth}
\centering
    \begin{tabular}{ccc}
    \hline
          \textbf{female} & \textbf{male} & \textbf{non-binary}\\\hline 
         7933 &17070 & 191  \\ \hline
    \end{tabular}
\end{subtable}
    \vfill
    \vspace{0.3em}
\begin{subtable}{\linewidth}
\centering
    \begin{tabular}{cccc}
    \hline
          \textbf{20-34} & \textbf{35-49} & \textbf{50-64} & \textbf{65+}\\\hline 
         9102 &4267 & 1739&1543 \\ \hline
    \end{tabular}
\end{subtable}
    \vfill
    \vspace{0.3em}
\begin{subtable}{\linewidth}
\centering
    \begin{tabular}{ccccc}
    \hline
          \textbf{pop} & \textbf{hip hop} & \textbf{rock} & \textbf{alternative}&\textbf{electronic}\\\hline 
         8863 &5277 & 2151&1735&1160  \\ \hline
    \end{tabular}
\end{subtable}
    \vfill
    \vspace{0.3em}
\begin{subtable}{\linewidth}
\centering
\setlength\tabcolsep{2.0pt}
    \begin{tabular}{ccccc}
    \hline
          \textbf{English} & \textbf{French} & \textbf{Italian} & \textbf{Portuguese}&\textbf{Japanese}\\\hline 
         11422 &4555 & 1978&1570&1429  \\ \hline
    \end{tabular}
\end{subtable}
    \vfill
    \vspace{0.3em}
\caption{Number of tracks for three genders, 4 age groups, 5 main genres and 5 main languages}
\label{genre}
\end{table}
Thus, automatic-strada can be used for various singing-voice-related tasks' training: singer sex classification, singer recognition, singer identification and singer age detection.

Moreover, for certain tasks mentioned above, having a dataset balanced across various subgroups and manually annotated could be beneficial for evaluation and bias analysis. Hence, we publish an annotated-strada consisting of 1200 segments from 200 tracks balanced across two genders (all singers are cis-gender), four age groups, and five languages. 

STraDa offers an unique opportunity for model training and bias analysis in singing-voice-related tasks thanks to its rich metadata, large-scale single lead singer and downloadable audio files. Therefore, we demonstrate the effectiveness of STraDa by fine-tuning a x-vector model for singer sex classification (SSC)\cite{snyder2018x}. Additionally, the annotated-strada enables us to conduct bias analysis across subgroups. Our findings highlight STraDa's utility for both enhancing model performance and facilitating bias analysis in singing voice related tasks.

\section{Related Work}
\subsection{Existing singer traits dataset}
\label{lr:datasets}

Publicly available datasets containing raw audio data spanning various genres, languages, and types of singers are valuable but rare. Datasets that include annotations detailing singer traits and track information are even scarcer. Moreover, the lead singer is often not identified, which leads to errors in the annotations in cases where there are multiple lead singers. Current datasets that contain singer metadata include the Jamendo4 dataset~\cite{ramona2008vocal}, which consists of 93 tracks, and the cante100 dataset~\cite{bayle2016large}, which includes 100 flamenco tracks. However, these datasets only cover limited amounts of genre and language of tracks, and we could not find age annotation. The public Karaoke1k dataset provides 1000 tracks and their cover tracks with singer metadata~\cite{bayle2017kara1k} for cover song detection, but it lacks age annotation and the identified lead singer, and the dataset is limited in genre and quantity. The public Jukebox dataset comprises 7000 excerpts of tracks from 936 different singers with singer gender~\cite{chowdhury2020jukebox}, but some tracks have multiple singers that are not annotated, which can result in incorrect annotations. 

In summary, there is no existing dataset that is large, diverse, and contains downloadable audio files as well as identified unique lead singer and cross-validated singer metadata all together.

\subsection{Singing voice related tasks}
A singer traits dataset could be beneficial for various tasks that suffer from the lack of diverse data. In the study of identifying singer traits, there has been limited research focusing on traits beyond distinguishing between sexes. Weninger et al. trained models to identify singer's sex, race, age and height, obtained an accuracy of 89.6\% for SSC on beat level using a Bidirectional Long Short-Term Memory (BLSTM), however the dataset only contains pop tracks that have quarter beats~\cite{weninger2011automatic}. Shi trained a model for sex classification and age detection that achieved and performance of 91\% and 36\% respectively using an internal dataset\cite{shi2015singer}. Jitendra et al. reported an accuracy of 89.16\% of SSC on 250 audio clips of only Indian music and an accuracy of 94.25\% on only 6 artists of dataset Artist20 ~\cite{jitendra2021singer}. While the diversity and quantity of data increase, the accuracy reduces to 70\% and and below 50\% respectively in~\cite{alonso2020tensorflow, bayle2017kara1k}. For the task of singer identification, researchers benchmark on the dataset Artist20 \footnote{\hyperlink{http://labrosa.ee.columbia.edu/projects/artistid/}{http://labrosa.ee.columbia.edu/projects/artistid/}} that contains only 20 singers where certain songs contain multiple singers and is limited in language and genre\cite{kim2021learning}. 

On a different note, previous studies often used the term \textit{singer gender classification}, but we argue that the term \textit{sex} is more suitable. This is because not only previous research relies on physiological distinctions, but also gender is recognized as a spectrum rather than a binary concept \cite{pinney2023much}.

\section{STraDa creation}

\subsection{Data sources of automatic-strada}
The data utilized to create automatic-strada was obtained from 4 distinct data sources: a music streaming service (Deezer), MusicBrainz (MB), Wikidata (WD) and Discogs (DC). MusicBrainz, Discogs, and Wikidata are all open music encyclopedias that gather music metadata and make it publicly available. 

Table \ref{info} displays the relevant information from these four data sources for each track and each artist that was utilized in the creation of the automatic-strada. The role column refers to the specific role of an artist in a given track, such as singer, composer, pianist, or guitarist.
\begin{table}[h!]
\hspace*{0.8cm}
\begin{tabular}{ccccccc}

\hline
 & \multicolumn{3}{l}{\textbf{For each track}}    & \multicolumn{3}{l}{\textbf{For each artist}}    \\ \hline
   & \multicolumn{1}{l}{\textbf{A}} & \multicolumn{1}{l}{\textbf{R}} & \multicolumn{1}{l}{\textbf{RD}} & \multicolumn{1}{l}{\textbf{G}} & \multicolumn{1}{l}{\textbf{BY}} & \multicolumn{1}{l}{\textbf{AC}}  \\ 
 \textbf{Deezer} & \multicolumn{1}{l}{\cmark} & \multicolumn{1}{l}{\cmark} & \multicolumn{1}{l}{\cmark} & \multicolumn{1}{l}{\cmark} & \multicolumn{1}{l}{\cmark} & \multicolumn{1}{l}{\cmark}  \\ 
\textbf{MB} & \multicolumn{1}{l}{\cmark} & \multicolumn{1}{l}{\cmark} & \multicolumn{1}{l}{\xmark} & \multicolumn{1}{l}{\cmark} & \multicolumn{1}{l}{\cmark} & \multicolumn{1}{l}{\cmark}  \\ 
 \textbf{WD} & \multicolumn{1}{l}{\cmark} & \multicolumn{1}{l}{\xmark} & \multicolumn{1}{l}{\xmark} & \multicolumn{1}{l}{\cmark}  &\multicolumn{1}{l}{\cmark} & \multicolumn{1}{l}{\cmark}  \\ 
 \textbf{DC} & \multicolumn{1}{l}{\cmark} & \multicolumn{1}{l}{\cmark} & \multicolumn{1}{l
}{\cmark} & \multicolumn{1}{l}{\xmark} & \multicolumn{1}{l}{\xmark} & \multicolumn{1}{l}{\xmark}  \\ \hline
\end{tabular}
\begin{tablenotes}
    \small
    \centering
    \item A: Artist, R: Role, RD: Release Date, G: Gender, \\BY: Birth Year, AC: Active Country
\end{tablenotes}
\caption{Relevant information in four data sources}
\label{info}
\end{table}

\subsection{Data matching and processing of automatic-strada}
We matched tracks and artists from four data sources using album names, track names, and singer metadata. To ensure only tracks with a single lead singer were included, we used \textit{role} information to identify singers with clear vocal-related roles, such as \textit{singer} or \textit{vocalist}, and removed tracks with multiple vocals. Moreover, we kept only audio segments with publicly accessible components and extracted voice-containing parts using lyrics alignment. In total, we obtained 25194 excerpts, each from a different song. For the \textit{release date} information, we opted for the earliest one among all data sources to align the data as closely as possible with the actual recording date of the song, which we consider to be more valuable data.

STraDa presents a list of annotations of track-related metadata and the corresponding lead vocalists. Notably, each track is uniquely associated with a sole designated lead singer, a unique characteristic that is not found in previous research.

STraDa's rich metadata supports diverse music-related research. Researchers have the opportunity to harness its capabilities for a range of applications. 

Furthermore, all music excerpts of 30 seconds, along with its metadata, are downloadable with a unique identifier from the music streaming service public API. 

\subsection{Annotated-strada's curation}

For evaluation on tasks such as singer age identification, singer sex classification and song language recognition, it is desirable to have a dataset with highly reliable labels to improve the reliability of the evaluation. Furthermore, obtaining balance in the distribution of testing data across various categories is important for a more thorough analysis of biases. Therefore, we completed STraDa by adding a manually verified testing dataset. 

We deliberately curated and annotated a dataset comprising 200 tracks of 200 singers that are evenly distributed across two sexes (female and male), five languages (English, French, Mandarin, Spanish, German), and four age groups of the singers (20-34, 35-49, 50-64, 65+). It's important to note that all singers chosen in the annotated-strada are cis-gender. This ensures that each language and each gender of the singer, within a specific age group, is represented by an equal number of five tracks. We chose different languages than the 5 most represented languages in the automatic-strada because it is important to have languages that come from different linguistic families to show the robustness of models across different languages. The age groups are chosen according to a similar research of Doukhan et al. in speech\cite{doukhan2018open}. From each track, we then selected 6 voice-containing segments of three seconds, which leads to a total number of 1200 segments. This amount is large enough to evaluate systems and conduct bias analysis for each subgroups. YouTube links and timestamps to all 1200 segments are provided for downloading, researchers could choose to use the whole track or segments of three seconds, depending on their own research need.

It's important to note that annotated-strada operates independently from automatic-strada. Annotated-strada can be downloaded using YouTube links, while automatic-strada is downloadable from the Deezer API. This is primarily due to the time-consuming nature of annotation. To obtain a sufficiently large dataset for evaluation and bias analysis, we lacked the resources to manually annotate 1200 tracks. Therefore, we opted to extract six segments from each track that varied in terms of timbre, accompaniment, and pitch. Additionally, excerpts in automatic-strada are limited to 30 seconds since they are preview sections of audio tracks in Deezer. It's nearly impossible to find six voice-containing segments that are sufficiently diverse within a 30-second excerpt. Hence, we chose to provide downloadable full-length tracks with timestamps to extract segments, rather than annotating 1200 excerpts from automatic-strada, which would have been much more time-consuming. While we acknowledge that this approach may not be as ideal as annotating all 1200 tracks, it represents a reasonable compromise given our resource constraints.

 Most of the tracks within the annotated-strada are copyrighted commercial tracks owned by various labels, which are more often listened to than royalty-free music. Consequently, we are unable to share downloadable audio files directly. We acknowledge that these links may not be stable or permanent, but they currently represent the most viable option for ensuring reproducibility. The collection and annotation process for the annotated-strada are conducted manually rather than through automated means to ensure accuracy. Future researchers are encouraged to consider annotated-strada as a potential dataset for evaluation and automatic-strada for training.

\subsection{Data coverage and limitations}

Our dataset has two primary limitations. Firstly, it is important to acknowledge that the representation of non-binary gender artists within automatic-strada is significantly lower than figures reported in non-binary gender studies~\cite{richards2016non}. Moreover, the accuracy of non-binary gender annotations derived from the four data sources we gathered cannot be definitively guaranteed. This is due to the inadequacy of information regarding the data collection procedures within these data sources. 

Secondly, STraDa has an overrepresentation of Western pop culture, a characteristic commonly observed in datasets encompassing cultural elements~\cite{wang2019gender}. Despite our effort to be inclusive of a variety of languages and genres, unlike datasets aforementioned in \autoref{lr:datasets}, it is important to acknowledge that the biases are still inevitable due to the overrepresentation of Western pop music within the data sources we used. 

As we continue to evolve this research, we wholeheartedly encourage researchers to contribute to STraDa.

\section{An use case: automatic SSC}
\label{automatic}
In this section, our objective is to show an use case of both automatic-strada and annotated-strada. The rich, extensive, and diverse features of automatic-strada make it an excellent dataset for training a system. The balanced and accurate features of annotated-strada render it an ideal dataset for evaluating a system's performance and conducting bias analysis.

The most complete metadata in automatic-strada is the gender information. Furthermore, in annotated-strada, all singers are cis-gender, indicating alignment between their gender identity and biological sex. Thus, a binary sex classification is suitable to be evaluated on annotated-strada. Considering that gender is more of a continuum rather than a binary construct, and is intertwined with internal self-perceptions and external societal perspectives on individuals~\cite{pinney2023much}, we decided to benchmark the task singer sex classification.

It is noteworthy that in automatic-strada, gender information is provided, but not biological sex. As a result, we excluded singers identifying as non-binary for this specific use case. Additionally, the majority of the population has congruent gender identity and biological sex\cite{goodman2019size}, which means in automatic-strada as well. We acknowledge there are mistaken cases where the singer's gender differs from their sex, but in this specific case, while testing on only cis-gender singers, we decided to use gender information in automatic-strada when training for the SSC system.

Here, our emphasis lies not solely on the system's performance, but rather on the comparative analysis of various algorithms and the exploration of biases present within the automated detection system by using annotated-strada.

\subsection{Baselines}
In this section, we present various baseline methodologies for sex detection. We explored several models such as a K-nearest neighbour classifier (kNN) along with fundamental frequency (f0), MFCCs and multilayer perceptrons (MLP), as well as a Convolutional Neural Network (CNN) and two variations of the x-vector system.

\subsubsection{F0 and kNN classifier (B1)}

Females produce approximately twice the male f0~\cite{hillenbrand2009role}. We extracted f0 from each excerpt in the automatic-strada by employing the full-capacity model of \textit{crepe} algorithm~\cite{kim2018crepe} and calculated an histogram of f0 for each excerpt. Then, the kNN algorithm was used to compute the similarity between the f0 histogram of the testing segment and the entire repository of training samples. We selected the five nearest neighbors from the training samples and assigned the sex label to the testing segment based on the majority of the five training samples.

\subsubsection{MFCCs and MLP classifier (B2)}

MFCCs are employed with SVM for speaker recognition in previous work~\cite{chen2009speaker,lee2012performance}. We used a MLP to classify the MFCCs features into two sex categories. Specifically, we extracted 13 MFCCs for separated voice of each excerpt from automatic-strada and annotated strada using the \textit{librosa} library~\cite{mcfee2015librosa}, and then implemented a four-layer MLP as the classifier for prediction. For voice separation, we used \textit{Spleeter}~\cite{spleeter2020}.

\subsubsection{CNN (B3)}
The CNN architecture employed is identical to the one proposed in a previous study of SSC~\cite{jitendra2021singer}. However, the authors used three-channel (RGB) spectrograms as input, whereas we opted to use a grey-scale spectrogram as input since it already encompasses all the necessary information. We extracted 3-second segments using separated voice or original as a data augmentation from the automatic-strada for training. We tested it on the original polyphonic music of annotated-strada.

\subsection{Fine tuning x-vector system on STraDa (X1\&X2)}
Given that the pre-trained x-vector system is trained on speech data, we propose to fine-tune the pre-trained x-vector system on automatic-strada. This approach allows us to adapt the model to better capture singing voices. Fine-tuning the x-vector system involves re-training the model on automatic-strada while retaining some of the pre-trained weights and architecture in the work of ~\cite{snyder2018x}. We reduced the embedding dimension from 512 to 64, and froze the first 3 layers to improve training efficiency. 

\subsubsection{Training Data}
During the training phase, we extracted three-second segments with voice present from automatic-strada. To reduce sex bias in the training data, we sampled twice as many segments from tracks performed by female singers as from those performed by male singers, since we have half the number of male singers as female singers. Finally, we obtained 177k three-second segments for training. 

\subsubsection{Data processing and augmentation}

In this study, we employed Mel-spectrograms, which are representations of sound signals in the frequency domain using 24 Mel filters, as the input for TDNN model to extract x-vector embeddings. To enhance the quality and quantity of our training data, we also incorporated source separation as a form of data augmentation. Specifically, for half of the training data, we utilized singing voices extracted from the excerpts instead of mixed tracks. This approach was intended to provide an alternative view of the song and encourage the model to focus more on the singing voice component. In order to compare the performance of the model with and without this data augmentation, we conducted experiments without (X1) and with (X2) data augmentation techniques.

\subsection{Results and analyses}
\subsubsection{Comparison of different systems}
\begin{figure}
    \centering
    \includegraphics[width=8cm]{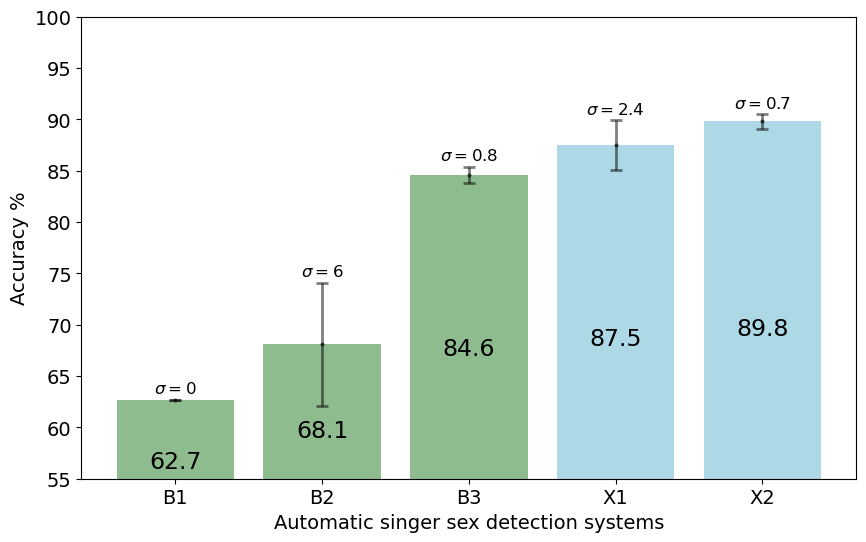}
    \caption{Accuracy (mean and standard deviation) of machine learning systems}
    \label{fig:acc}
\end{figure}

We used accuracy as the evaluation metric, which is calculated by dividing the number of true positives by the total number of samples. Standard deviation is calculated by training model with different initialization five times. Figure \ref{fig:acc} indicates that X1 and X2 demonstrate superior performance when compared to all baseline systems. Fine-tuning x-vector on automatic-strada with source separation as data augmentation yielded the highest accuracy. Across five experiments with different initial weights, this method achieved an average accuracy of 89.8\%. These results suggest that the use of both separated voices and original polyphonic music as data augmentation enhances the system's ability to handle background accompaniment and improves its robustness. These findings demonstrate that the fine-tuning x-vector model is capable of capturing distinct characteristics of male and female voices across diverse data.

\subsubsection{Model biases}
\label{machine biases}
The annotated-strada consists of 1200 segments balanced across gender, language, and age groups, which enables the comparison of performances on different subgroups. We used the metric of recall, which is defined as the ratio of correctly identified positive samples to the total number of samples in a given category. It shows how accurate the model is in predicting in a specific subgroup. Standard deviations are calculated across 5 experiments with different initial weights.

\begin{table}[H]
\centering
\begin{subtable}{\linewidth}
\centering
\begin{tabular}{ccc}
\hline
    \textbf{Female}  & \textbf{Male}  \\ \hline
    85.6$\pm$2.5 & \textbf{94.1 $\pm$1.2}  \\\hline
    \end{tabular}
    \end{subtable}
    \vfill
    \vspace{0.3em}
\begin{subtable}{\linewidth}
\centering
\begin{tabular}{cccc}
\hline
    \textbf{20-34} & \textbf{35-49} & \textbf{50-65} & \textbf{66+} \\ \hline
    91.3$\pm$1.5 & \textbf{93.6$\pm$1.0} & 86.0$\pm$1.0 & 88.3$\pm$2.0 \\\hline
    \end{tabular}
\end{subtable}
    \vfill
    \vspace{0.3em}
\begin{subtable}{\linewidth}
\centering
\setlength\tabcolsep{1.7pt}
\begin{tabular}{ccccc}
\hline
    \textbf{French} & \textbf{English} & \textbf{Mandarin}& \textbf{Spanish}& \textbf{German} \\ \hline
    87.4$\pm$0.6 & 88.8$\pm$2.5 & \textbf{93.9$\pm$0.8} & 89.2$\pm$0.4 & 90.0$\pm$1.9\\\hline
    \end{tabular}
\end{subtable}
\caption{Recalls (\%) of different subgroups of singers}
\label{recall}
\end{table}

Table \ref{recall} indicates that the model exhibits superior performance in recognizing male voices in comparison to female voices, a finding consistent with results reported in the work of automatic speaker gender detection by Doukhan et al.~\cite{doukhan2018open} and automatic speech recognition (ASR) by Garnerin et al.~\cite{garnerin2019gender}. Regarding age, the model exhibits a lower recall for singers older than 50 years old, a trend that has also been observed in studies reporting lower performance on older speakers in ASR ~\cite{feng2021quantifying, sawalha2013effects}. Furthermore, the x-vector model demonstrates a greater capacity for accurately detecting sex in Mandarin tracks as opposed to French tracks, which might be due to the high-pitched female voices in traditional Chinese folk music.

In this section, we demonstrated how automatic-strada enables us to train multiple SSC models using diverse songs spanning different genres and languages. The 1200 annotated segments in annotated-strada enables us to evaluate these models across a variety of songs. Additionally, due to its balance across subgroups, we were able to compare performances of subgroups on SSC. 

\section{Conclusion}
In this study, to provide the community a dataset with downloadable audio files and rich metadata, we released STraDa, with more than 25k tracks and 5k unique lead singers in automatic-strada, that contains rich track and singer metadata, and an annotated-strada which is balanced across different subgroups. We showed one of its potential use cases by benchmarking SSC on STraDa and compared performances in each subgroup. Moving forward, STraDa could be extended to other MIR tasks, especially singing-voice-related tasks, such as singer identification. 

Furthermore, there is still space for future work. We could enrich automatic-strada by adding more tracks from underrepresented groups, such as non-English pop songs and female and non-binary gender singers. We could also complete annotated-strada by annotating additional information, such as music genre, and adding tracks of non-binary singers. Additionally, research into the reasons behind model bias in SSC could also be conducted.

We extend a warm invitation to future researchers to use and enrich STraDa for a variety of tasks.

\bibliographystyle{IEEEtran}
\bibliography{mybib}

\end{document}